\documentclass[final, 12pt]{elsarticle}
\makeatletter
\def\ps@pprintTitle{%
 \let\@oddhead\@empty
 \let\@evenhead\@empty
 \def\@oddfoot{\centerline{\thepage}}%
 \let\@evenfoot\@oddfoot}
\makeatother
\pdfoutput=1 
\usepackage[main=english,slovak]{babel}
\usepackage[utf8]{inputenc}
\usepackage[T1]{fontenc}
\usepackage{flushend}
\usepackage{amsmath}
\usepackage{array}
\usepackage[noend]{algpseudocode}
\usepackage{algorithm}
\usepackage{url}
\usepackage{hyperref}

\usepackage{amssymb}

\newcounter{bla}

\journal{arXiv}

\begin{document}

\begin{frontmatter}



\title{Acceleration of Mean Square Distance Calculations with Floating Close Structure in Metadynamics Simulations}


\author[uvt]{Jana Paz\'urikov\'a\corref{mycorrespondingauthor}}
\ead{pazurikova@ics.muni.cz}

\author[uvt]{Jaroslav O\v{l}ha}
\author[uvt]{Ale\v{s} K\v{r}enek}
\author[vscht]{Vojt\v{e}ch Spiwok}

\cortext[mycorrespondingauthor]{Corresponding author}

\address[uvt]{Institute of Computer Science, Masaryk University, Botanick\'a 554/68a, 602 00 Brno, Czech Republic}
\address[vscht]{Department of Biochemistry and Microbiology, University of Chemistry and Technology, Technick\'a 5, 166 28 Prague 6, Czech Republic}

\begin{abstract}
Molecular dynamics simulates the~movements of atoms. Due to its high cost, many methods have been developed to ``push the~simulation forward''. One of them, metadynamics, can hasten the~molecular dynamics with the~help of variables describing the~simulated process. However, the~evaluation of these variables can include numerous mean square distance calculations that introduce substantial computational demands, thus jeopardize the~benefit of the~approach. Recently, we proposed an~approximative method that significantly reduces the~number of these distance calculations. Here we evaluate the~performance and the~scalability on two molecular systems. We assess the~maximal theoretical speed-up based on the reduction of distance computations and Ahmdal's law and compare it to the~practical speed-up achieved with our implementation. 
\end{abstract}

\begin{keyword}
approximation \sep acceleration \sep molecular dynamics simulation \sep metadynamics
\end{keyword}

\end{frontmatter}

\section{Introduction}

Molecular dynamics~\cite{Jensen2007} is a~well-known computational method which simulates
the behavior of a~molecule at the~atomic scale by integrating Newton's equations of motion for all atoms over time.
Due to a~high frequency of movements at this scale, the~integration step must not exceed femtoseconds, which yields realistic computational speed to be at most nanoseconds
of simulated time per day of computation time.
On the~other hand, biologically relevant phenomena take milliseconds and more at the
macromolecular level, hence their complete simulations are still not feasible nowadays.
The necessity of long simulation timescales can be traced down to high energy barriers between the~distinct states of the~macromolecule, which require a~considerable momentum to be gained at the~nanoscale before they are overcome.

\begin{figure}[h]
\begin{center}
\includegraphics[width=.8\hsize]{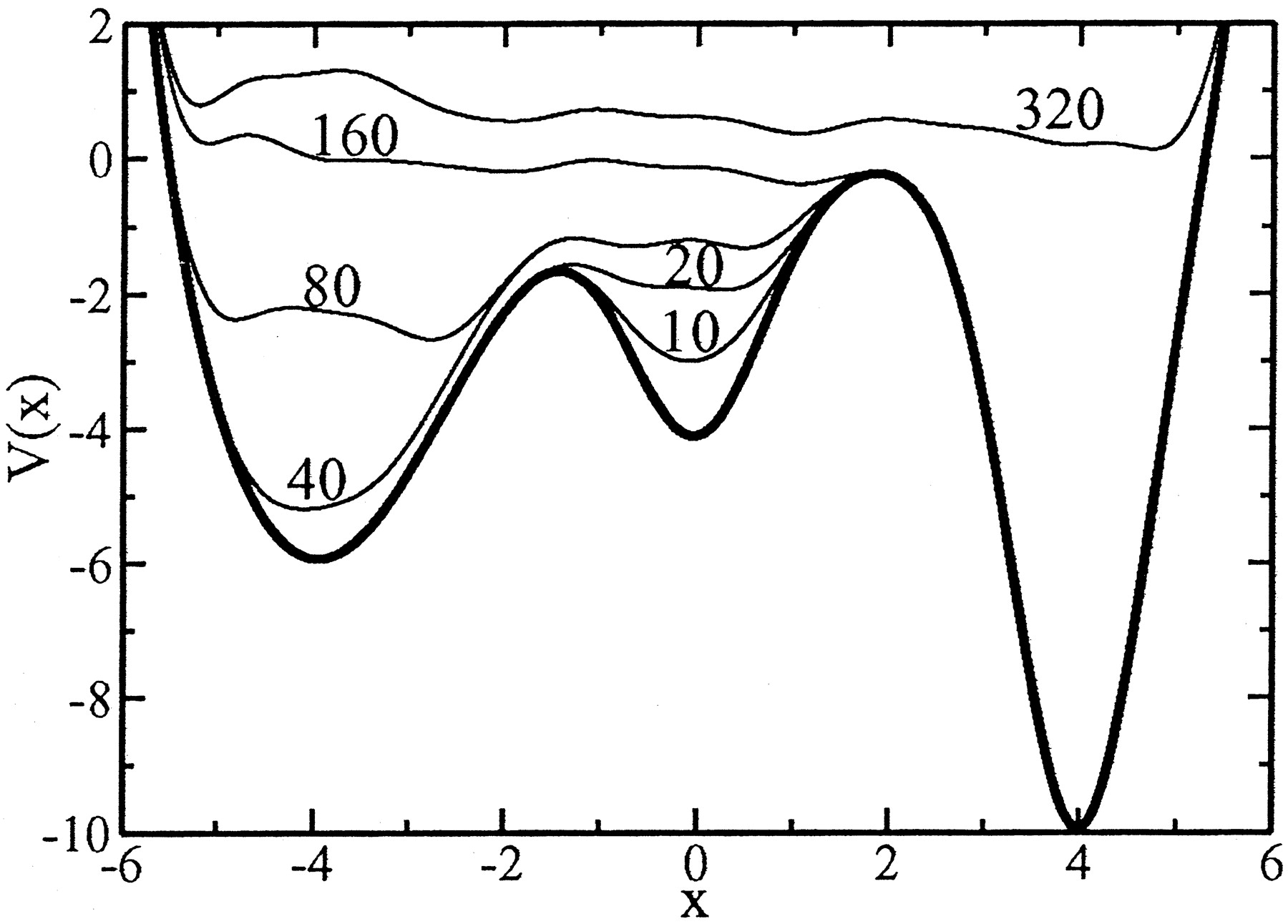}
\end{center}
\caption{Filling potential energy minima with a~bias potential. The simulation starts with $x=0$, and it accumulates the~bias potential in this area first. After 20 steps the~energy barrier at $x=-1.75$ is overcome and the~bias potential starts filling the~minimum at $x=-4$. After approx.\ 300 steps the~state space is explored. Taken from \cite{Laio2002}, image copyright (2002) National Academy of Sciences.}
\label{f:sand}
\end{figure}

However, various classes of problems do not require a~realistic trace of the~
simulation, it is sufficient to explore the~state space in which the
simulation would travel eventually~\cite{Bernardi2015}.
Therefore, the~simulation run can be ``pushed forward'' by additional steering,
e.g.\ in the~method called \emph{metadynamics} by adding \emph{bias potential} 
in already visited areas, or by ``filling energy minima with computational sand'', as illustrated in Figure~\ref{f:sand}.

Many methods have been developed to accelerate molecular dynamics simulations by artificial intervensions~\cite{Tiwary2016}.
Unfortunately, the~most promising ones often introduce a~non-negligible 
computational overhead to calculate the~bias potential, hence jeopardize the~benefit
of the~whole approach.
Recently, we proposed a~method~\cite{Pazurikova2017} which considerably reduces the~number
of required calls to the~most expensive routine of such calculation---\emph{mean square distance} (MSD) between two molecular shapes, while not affecting the~overall accuracy.
Here we introduce the ready-to-use implementation of the~method (replacing a rather inefficient prototype from \cite{Pazurikova2017}) and evaluate the~practical benefits---speed-up of the real computation, including the~use of multiple CPU cores.

\section{Background}

Simulations of molecular dynamics (MD) calculate the~movements of atoms caused by their interactions \cite{Jensen2007}. First, the~potential energy of the~system has to be computed. In this work, we modeled the~interatomic potential using molecular mechanics model, considering atoms as mass charged points and approximating both bonded and nonbonded interactions with a~series of empirical functions and parameters. Second, the~forces (determined by the~potential) are integrated with Newton's law of motion to obtain the~positions of atoms in the~next step. 

Computational demands of molecular dynamics grow with the~number of atoms and simulation steps. The calculation of the~interatomic potential scales as $\mathcal{O}(N_{atoms}\log(N_{atoms}))$ with common methods. The parallelization through the~spatial domain makes it possible to simulate even millions of atoms. On the~other hand, the~timescales of simulations are more difficult to handle. A~tiny timestep of the~integration scheme (in the~order of $10^{-15}$\,s) makes it hardly feasible to simulate more than a~few microseconds \cite{Durrant2016}, a millisecond at most \cite{Lindorff2016}, although many biologically or chemically relevant processes take longer than that. We need longer simulation times because we need to explore the~whole energy landscape, ideally all possible states of the~molecular system. However, molecules tend to stay in the~local energy minima and they cross the~high-energy barrier between them only once in a~while. 

Many methods have been developed to accelerate the~rare events of those crossings, for an overview see \cite{Spiwok2015b, Tiwary2016, Laio2008}. They enhance the~sampling of the~energy landscape and explore it much faster using various artificial interventions to the~simulation. One of the~most common, metadynamics \cite{Laio2002} (MTD), fills the~minima in the~energy landscape with an~artifical bias potential. It ``spreads the~sand into the~valleys'', and thus makes it easier to cross the~peaks.

To properly function, metadynamics needs a~few collective variables (CV)---preselected degrees of freedom that differentiate between the~states of the~simulated process and encompass all the~relevant motions. Moreover, they have to be limited in number, otherwise, both the~performance and the~accuracy of metadynamics will decline \cite{Laio2005, Barducci2011}.  Collective variables are usually computed from the~coordinates of the~molecular system at the~given time, further referred to as the~\emph{current structure} or $\mathbf{x}$. 

Metadynamics accumulates a~bias potential in the~form of Gaussian hills centered in the~values of collective variables, see Equation (\ref{eq:close:vbias}). In Figure~\ref{f:sand}, the~sum of Gaussian hills forms curvy lines depicting the~borders of the~summed bias potential in various time points. Well-chosen CVs direct to the~right spot for ``the sand to be spread on''. The bias potential expresses as the~bias force, so collective variables need to be differentiable with respect to coordinates.

 \begin{equation}  
  \begin{split}
     \label{eq:close:vbias}
   V_{bias}(\mathbf{S}(\mathbf{x}), t) = \sum_{\substack{t' = \tau_{G}, 2\tau_{G}, \dots \\ t < t'}} \prod_{i} w \exp{\left( -\frac{(S_i(\mathbf{x}(t)) - s_i^{t'})^2}{2\sigma^2} \right)} \\
   F_{bias}(t) = \frac{\partial V_{bias}(\mathbf{S}(\mathbf{x}), t)}{\partial \mathbf{x}}\\   
  \end{split}
  \end{equation}
 \normalsize{where $\mathbf{x}$ are coordinates of the~current structure, $\mathbf{S}(\mathbf{x})$ is the~vector of functions computing collective variables; $\tau_{G}$ is the~frequency of Gaussian hills addition; $w$ and $\sigma$ is the~height and the~width of the~Gaussian hill; $s_i^{t'}$ is the~value of $i$-th CV at time $t'$, i.e. $S_i(\mathbf{x}(t'))$.}
  \normalsize

The choice or design of CVs requires physico-chemical expertise and naturally gets more complicated with complex processes. A distance between specific atoms, angles or their combinations work well for simple processes. More complex ones, like protein folding, require a~more sophisticated approach such as \emph{path CVs} \cite{Branduardi2007} or their generalization, \emph{property map CVs} \cite{Spiwok2011}. They are based on comparison to reference landmark structures, a~series of snapshots capturing various states of the~process. Reference structures can be generated with any non-continuous simulations, non-physical actions, high temperatures, annealing techniques or even manual construction in a~visualization software. 

The value of the~property map collective variable corresponds to a weighted average of properties of the~closest reference structures, where the~weight is determined by the~distance, see Equation (\ref{eq:close:property_map_cv}). For example, if we take an~index of the~reference structure as a~property, the~value of CV would be close to the~index of the~closest reference structure.


\begin{equation}
   S(\mathbf{x}) = \frac{\sum\limits_{i=1}^{N} q_i \exp{\left(-\lambda D(\mathbf{x}, \mathbf{a}_i)\right)}}{\sum\limits_{i=1}^{N} \exp{\left(-\lambda D(\mathbf{x}, \mathbf{a}_i)\right)}}
   \label{eq:close:property_map_cv}
  \end{equation}
  \normalsize{where $N$ is the~number of reference structures, $\mathbf{a}_i$ are coordinates of $i$-th reference structure, $q_i$ is an~arbitrary property of the~given reference structure, $D()$ is the~distance function, $\lambda$ is a~tuning parameter.}  
  \normalsize

Clearly, the~evaluation of property map CVs requires as many distance computations in every step of the~simulation as there are reference structures. For smooth processes, we can reduce the~number of distance computations by employing a~\emph{neighbourlist} (NL). Since the~contribution to a~CV's value decreases exponentially with the~distance, without sudden changes in system's structure it suffices to compute distance only to the~few closest reference structures in the~NL and update the~list regularly. However, in processes with abrupt changes, we need to compute the~distances of the~current structure to all the~reference structures in each step (referred to as \emph{no neighbourlist}). In both cases, numerous distance computations introduce high computational cost.

\section{Floating Close Structure}

The standard distance measure between molecular systems in computational sciences is the~(root) mean square distance between superimposed, fitted structures \cite[ch.16]{Gu2009}. The fitting between structures minimizes the~distance and consists of two tasks: translation and rotation. Shifting both structures to the~same location in the~coordinate system means subtracting coordinates of the~center of mass from coordinates of all atoms. As this is both simple and fast, we omit it from further equations for simplicity and assume that it was done in advance for all the~structures. On the~contrary, obtaining the~rotation matrix presents the~most expensive task in the~distance computation. It involves demanding matrix operations, such as the matrix diagonalization from quaternions as in \cite{Kearsley1989, Coutsias2004} or the singular value decomposition as in \cite{Kabsch1976}. 

To get a~closer look on the~bottlenecks, we analyzed the~performance of simulations with a profiling software. We ran simulations using Gromacs \cite{Pall2015} as the~molecular dynamics software and Plumed \cite{Tribello2014} as the~metadynamics tool, details in section \ref{sec:dataset}. Even with the~neighbourlist employed, distance computation takes up to 43\% of miniprotein Trp-cage simulation walltime. For small 24-atom cyclooctane simulation without the~neighbourlist, it takes 93\% of walltime, see Table~\ref{tab:ahmdal}. The majority of that time is spent in BLAS function DSYEVR, diagonalizing matrix in Kearsley's method \cite{Kearsley1989} implemented in Plumed. As BLAS is already a~finely-tuned routine, we assume no further code optimizations would bring a~substantial improvement. Offloading to GPU brings its own issues, as we outlined in \cite{Filipovic2016}. Therefore, we aim to modify the~entire method in a~way that reduces the~number of demanding distance computations and thus decreases 
the~computational cost.

\subsection{Exact MSD Calculation}
\begin{figure}[h]
 \centering
 \includegraphics[width=0.8\hsize]{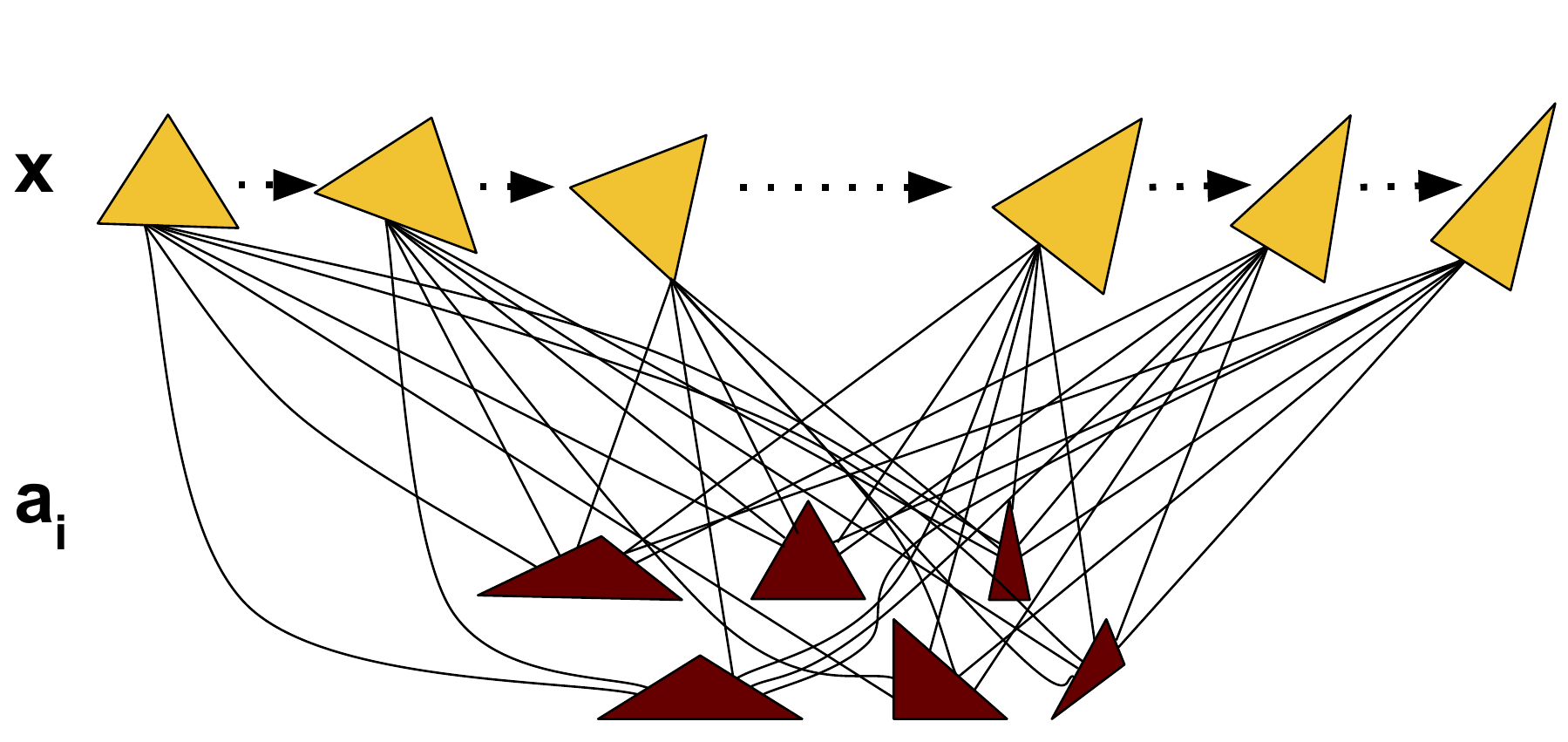}
 \caption{Original method expensively computes distances from the current structure ($\mathbf{x}$) to reference structures in neighbourlist ($\mathbf{a}_i$) in every step. Arrows denote passing simulation time.} \label{fig:diagram-original}
\end{figure}

 The original method for exact MSD computation calculates accurately but expensively the~rotation matrix for all the~reference structures in the~neighbourlist in every step, see Algorithm~\ref{alg:model} and Figure~\ref{fig:diagram-original}. Then it uses rotation matrices to compute the~distance, see Equation (\ref{eq:dist}), and derivatives with respect to coordinates (to evaluate the~bias force), see Equation (\ref{eq:der_dist}).

\begin{algorithm}
 \caption{Computation of the~collective variables in the~original method} \label{alg:model}
 \ttfamily
\begin{algorithmic}[1]
 \Function{compute\_cvs}{$\mathbf{x}$, ref\_structures}
 \Loop\ through $\mathbf{a_i}$ in neighlist
    \State $R_{\mathbf{x}\mathbf{a}_i} \gets$ compute\_R($\mathbf{x}$, $\mathbf{a}_i$)
    \State distance $\gets$ Equation (\ref{eq:dist})
    \State derivatives $\gets$ Equation (\ref{eq:der_dist})
 \EndLoop
 \EndFunction
 \end{algorithmic}
\vspace{1em}
 \end{algorithm}
 \rmfamily

 \begin{equation}
   \begin{split}
    D(\mathbf{x}, \mathbf{a}_i) = \sum_j^{\mathbf{N}_{atoms}}w_j \|\mathbf{d}_j\|^2 \\
    \text{where $\mathbf{d}_j = \mathbf{x}_j  - R_{\mathbf{a}_i\mathbf{x}}\mathbf{a}_{ij}$}
    \label{eq:dist}
   \end{split}
  \end{equation}
  \normalsize
  \begin{equation}
   \begin{split}
    \frac{\partial D(\mathbf{x}, \mathbf{a}_i)}{\partial \mathbf{x}_k} = 2w_k\mathbf{d}_k + w'_k \left(\sum_j{{-2}w_j \mathbf{d}_j}\right) + \\ \left( \sum_j -2 w_j \mathbf{d}_j \times \mathbf{a}_{ij} \right) \otimes \frac{\partial  R_{\mathbf{a}_i\mathbf{x}}}{\partial \mathbf{x}_k}\\
    \label{eq:der_dist}
   \end{split}
  \end{equation}
  \normalsize{where $\mathbf{x}$ are coordinates of the~current structure, $\mathbf{a}_i$ are coordinates of $i$-th reference structure, $R_{\mathbf{a}_i\mathbf{x}}$ is the~rotation matrix fitting the~reference structure $\mathbf{a}_i$ onto the~current structure $\mathbf{x}$, $w$ and $w'$ are weights, $\otimes$ is element-wise multiplication.}
  \normalsize
  
 Two sets of weights influence the~computation: \emph{displacement} $w$ that describes the~contribution of an~atom's displacement to the~total distance, and \emph{alignment} $w'$ that describes the~contribution to the~center of mass which influences the~translation. 
  
 During a~usual step, we need to compute the~expensive rotation matrix $M$ times, where $M$ is the~neighbourlist size. During an~update of NL, every $L$ steps, we calculate the~distance (thus also the~rotation matrix) to all the~reference structures. More details about the~computational cost in section \ref{sec:theory}.
  
\subsection{Approximative MSD Calculation}


Our method is based on the~assumption that the~molecular system moves reasonably smoothly in dynamics simulations due to a~tiny timestep. As its coordinates alter only slowly, the~values in the~rotation matrices should also change only slightly every step. Therefore, we reuse them in a~few consecutive steps, until the~current structure moves too far. 

\begin{figure}[h]
\centering
 \includegraphics[width=0.8\hsize]{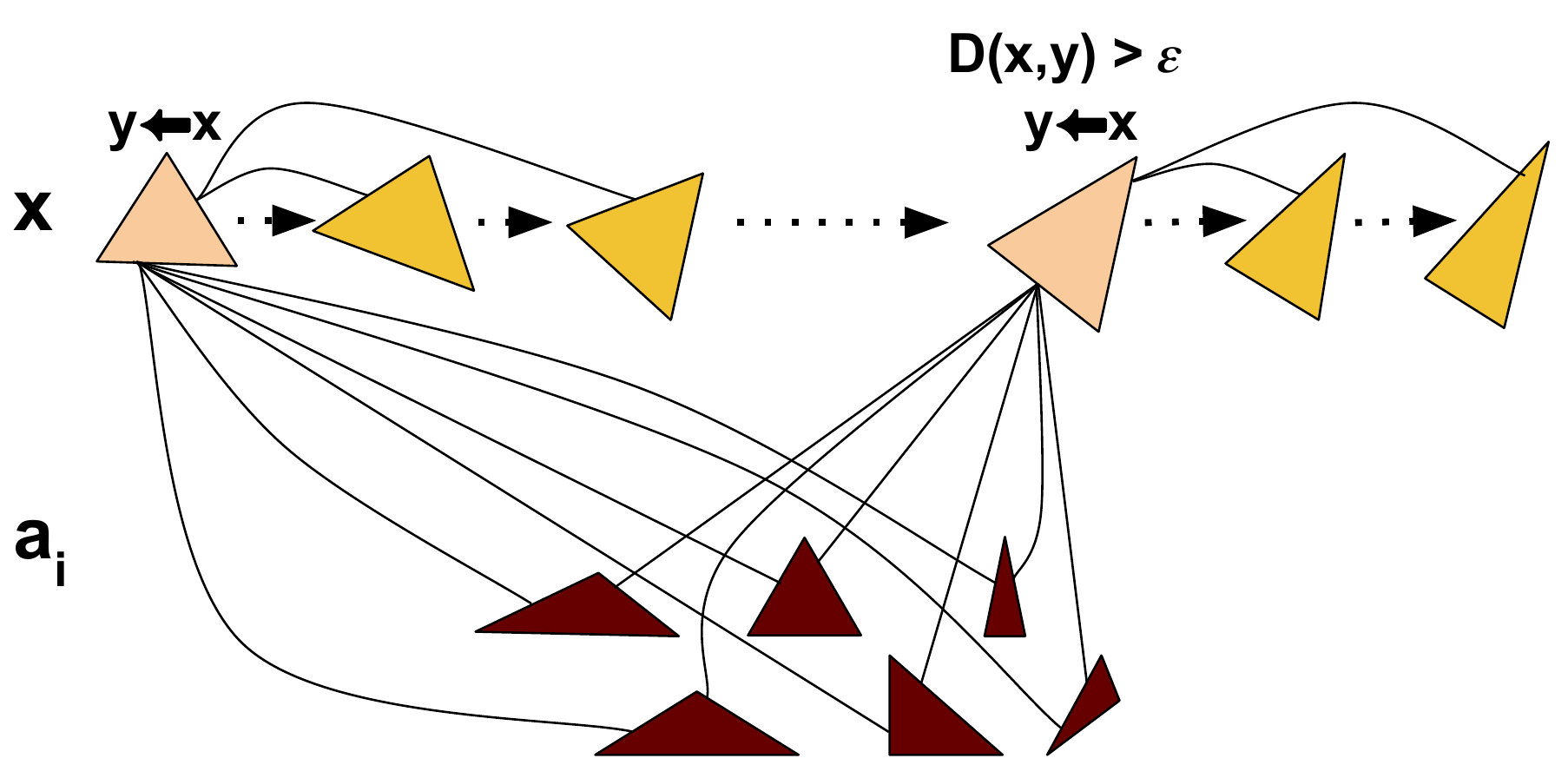}
 \caption{Close structure method expensively computes distances only between the~current ($\mathbf{x}$) and the~close structure ($\mathbf{y}$) and then during reassignments when the distances to all reference structures ($\mathbf{a}_i$) are accurately calculated. Arrows denote passing simulation time.} \label{fig:diagram-close}
\end{figure}

At the~beginning of the~simulation, we start as usual, see Figure~\ref{fig:diagram-close} and Algorithm~\ref{alg:approx_model}. We assign current coordinates to a~\emph{close structure} $\mathbf{y}$ and save computed rotation matrices $R_{\mathbf{a}_i\mathbf{y}}$. In most steps, we reuse these saved matrices to approximate $R_{\mathbf{a}_i\mathbf{x}} \approx R_{\mathbf{x}\mathbf{y}}R_{\mathbf{a}_i\mathbf{y}}$ while the~current structure $\mathbf{x}$ and the~close structure $\mathbf{y}$ stay close to each other. If the~distance $D(\mathbf{x}, \mathbf{y})$ exceeds a~given threshold $\varepsilon$, we reassign the~close structure, $\mathbf{y} \gets \mathbf{x}$ and recompute the~matrices $R_{\mathbf{a}_i\mathbf{y}}$.

 \begin{algorithm}
 \caption{Approximative computation of the~collective variables with the~close structure} \label{alg:approx_model}
 \ttfamily
\begin{algorithmic}[1]
 \Function{compute\_cvs}{$\mathbf{x}$, ref\_structures}
 \State $R_{\mathbf{x}\mathbf{y}} \gets$ compute\_R($\mathbf{x}$,$\mathbf{y}$) 
 \State $D_{\mathbf{x}\mathbf{y}} \gets$ Equation (\ref{eq:dist})
 \State $\partial R_{\mathbf{x}\mathbf{y}}/\partial \mathbf{x}_k \gets$ Equation(\ref{eq:der_dist})
 \If {$D_{\mathbf{x}\mathbf{y}} > \varepsilon$}

    \Loop\ through all $\mathbf{a_i}$
      \State $R_{\mathbf{a}_i\mathbf{x}} \gets$ compute\_R($\mathbf{x}$, $\mathbf{a}_i$)
      \State save $R_{\mathbf{a}_i\mathbf{x}}$
      \State distance $\gets$ Equation (\ref{eq:dist})
      \State derivatives $\gets$ Equation (\ref{eq:der_dist})
    \EndLoop
 \Else
    \Loop\ through $\mathbf{a_i}$ in neighlist
       \State $R_{\mathbf{a}_i\mathbf{x}} \gets R_{\mathbf{x}\mathbf{y}}R_{\mathbf{a}_i\mathbf{y}}$
       \State distance $\gets$ Equation (\ref{eq:close_dist})
       \State derivatives $\gets$ Equation (\ref{eq:close_der_dist})
    \EndLoop
 \EndIf 
 \EndFunction
 \end{algorithmic}
\vspace{1em}
 \end{algorithm}
 \rmfamily

We use exact MSD calculation, Equations (\ref{eq:dist}) and (\ref{eq:der_dist}), in the~reassignment steps and for the~computation of $D(\mathbf{x}, \mathbf{y})$. In the~majority of steps, we use Equation (\ref{eq:close_dist}) for distance and Equation (\ref{eq:close_der_dist}) for derivatives. $R_{\mathbf{x}\mathbf{y}} R_{\mathbf{a}_i\mathbf{y}}$ approximates $R_{\mathbf{a}_i\mathbf{x}}$, omitting the~demanding accurate computation of the~rotation matrix. 

 \begin{equation}
  \begin{split}
   \tilde{D}(\mathbf{x}, \mathbf{a}_i) = \sum_j^{\mathbf{N}_{atoms}} w_j \| \tilde{\mathbf{d}_j} \| ^2 \\
   \tilde{\mathbf{d}_j} = \mathbf{x}_j  - R_{\mathbf{x}\mathbf{y}} R_{\mathbf{a}_i\mathbf{y}} \mathbf{a}_{ij}
 \label{eq:close_dist}
  \end{split}
 \end{equation}

\begin{equation}
\begin{split}
 \frac{\partial \tilde{D}(\mathbf{x}, \mathbf{a}_i)}{\partial \mathbf{x}_k} = 
2w_k\tilde{\mathbf{d}_k} + w'_k \left( \sum_j {-2}w_j \tilde{\mathbf{d}_j} \right) +  \\
\left( \sum_j -2 w_j \tilde{\mathbf{d}_j} \times R_{\mathbf{a}_i\mathbf{y}} \mathbf{a}_{ij} \right) \otimes \frac{\partial R_{\mathbf{x}\mathbf{y}}}{\partial \mathbf{x}_k}
 \label{eq:close_der_dist}
 \end{split}
\end{equation}
where $\mathbf{y}$ is the close structure, $R_{\mathbf{x}\mathbf{y}}$ is the rotation matrix that fits the close structure to the current structure, $R_{\mathbf{a}_i\mathbf{y}}$ is the saved rotation matrix that fits the close structure to the reference structure $\mathbf{a}_i$.
 
The neighbourlist technique, not detailed in Algorithm~\ref{alg:approx_model} for simplicity, can be applied as usual in conjunction with the~close structure method. However, we do not recompute the~distance to all reference structures in the~NL update step, as it is in the~original method. Instead, we recompute during the~reassignment of the~close structure. Therefore, in the~close structure method, we select the~structures during the~NL update according to the~approximated distance.

Close structure approximation reduces the~number of distance computations in a~usual step to one---evaluating $D(\mathbf{x}, \mathbf{y})$. When reassigning the~close structure, we expensively calculate the~distance to all $N$ reference structures. More details about the~computational cost in section \ref{sec:theory}.

\section{Implementation}

We implemented the~close structure method into Plumed \cite{Tribello2014}, a~standard tool for metadynamics. Plumed works as a~plugin into many molecular dynamics softwares, we used it with Gromacs \cite{Pall2015}. We changed two classes, one that represents property map and path CVs, and another that encompasses RMSD computation. Since the prototype implementation of our method, we have dealt with issues regarding code optimization, data structures, memory access patterns, communication through MPI and scalability issues. The code ready for production use is available in the version v2.4 of Plumed, see \url{https://github.com/plumed/plumed2}. 

For performance evaluation, we combined Gromacs v5.1.4 with Plumed v2.3. We applied the~same level of minor code optimizations, enforcing the~vectorization of loops and inlining of template functions, both on our modified and the~original code in parts regarding the~evaluation of property map collective variables. We did not include any aggressive optimizations that would transform data structures since such heavy changes to software design would make it difficult to incorporate the~code into the official software \cite{plumed-code}. We compiled the~code with Intel compiler version 17 and SIMD instruction set AVX\_256.

\section{Computational Details}
\label{sec:dataset}
In this paper, we focus on the~evaluation of the acceleration and the scalability. We reuse two molecular systems and their simulation setup from \cite{Pazurikova2017} where we have thoroughly evaluated the~accuracy. Thus, we describe datasets only shortly, for further details see \cite{Pazurikova2017}.

The first molecular system, a~non-symmetric trans,trans-1,2,4-triflourocyclo\-octane (referred to as \emph{cyclooctane} throughout the~paper) can form different conformations with rapid transitions. As it contains only 24 atoms and is simulated in the vacuum (ergo no further molecules of the~solvent), the~usually high computational cost of molecular dynamics stays low for this molecule. However, abrupt and rare changes in structure require the~computation of distances to all 521 reference structures in every step, because the~neighbourlist would not be able to keep up. As distance computations take a~vast majority of the~computation time, it serves as an~example of a~molecule that should greatly benefit from the~acceleration of CV computation. 

Second, 304-atom miniprotein Trp-cage, PDB ID 1L2Y, simulated in an~implicit solvent smoothly diffuses between several conformations in its slow folding. The molecular dynamics, especially long-range electrostatic interactions, takes the~majority of the~computation time. Even with the~neighbourlist, the~closest 50 structures chosen from all 2120 reference structures every 50 steps, the~contribution of metadynamics is not negligible, it takes almost half of the~computation time. The trp-cage simulation shows a~more common example where the~acceleration of metadynamics speeds up the~whole simulation only modestly. However, reducing computational overhead brought by metadynamics enables us to simulate at the~speed of classical molecular dynamics with the~advantage of rare events' acceleration with metadynamics. 

We have set up the~simulation as described in \cite{Pazurikova2017}. All simulations ran for 100,000 steps. Metadynamics simulation has been done without a~neighbourlist for cyclooctane, i.e.distances to all 521 reference structures calculated every step, with 50-item NL updated every 50 steps for Trp-cage. In all simulations, the~bias potential has been computed with the~grid \cite{plumed-metad}, diminishing the~cost of a Gaussian hill summation growing with longer simulations. In close structure simulations, we reassign the~close structure if $D(\mathbf{x}, \mathbf{y})> \varepsilon$, where $\varepsilon= 0.01$\,nm$^2$. 

Experiments for this paper differ from those in \cite{Pazurikova2017} in versions of software and our method's implementation. Accuracy evaluation there has been done on Gromacs 4.5.7 and Plumed 2.1 with the prototype implementation of our method. Performance evaluation here has been done on Gromacs 5.1.4 and Plumed 2.3 with ready-to-use, optimized implementation of our method. All performance evaluation experiments were executed on one machine with 2x 8-core Intel Xeon E5-2650 v2 2.6 GHz, even multiple MPI processes were assigned to the~same machine.

\section{Results and Discussion}
The~main aim of the~close structure method is to accelerate metadynamics simulations by reducing the~number of distance computations. Therefore, in its evaluation, we focused both on the theoretical and the practical speed-up. We have evaluated the~accuracy of the~close structure method meticulously in \cite{Pazurikova2017}, here we only shortly outline the~main findings.

\subsection{Accuracy}
Because of the~chaotic character of molecular dynamics, even the~smallest changes caused by the~close structure approximation cause subsequent significant changes in trajectory. Since the~step-by-step comparison of trajectories would be meaningless, we inspected the~similarity of explored energy landscapes. First, we evaluated both Equation (\ref{eq:dist}) and (\ref{eq:close_dist}) on a~long series of $\mathbf{x}$ and corresponding $R_{\mathbf{x}\mathbf{y}}$, finding almost perfect correlation. Second, we visualized property map collective variables computed during the~simulation with the~original method and with the~close structure method. Both showed a great resemblance. Finally, we analyzed essential coordinates \cite{Amadei1993} of both trajectories. We found a~match not only between the~essential movements of the~system but also in minor harmonic oscillations. We concluded that the~trajectory of the~close structure simulation explored the~same energy landscape as the~simulation with the~original method. 

\subsection{Theoretical Speedup}
\label{sec:theory}
For theoretical evaluation of speed-up, we inspected the~number of expensive distance computations and conditions for speed-up.In the next section, we calculate the theoretical speed-up according to Ahmdal's law for both evaluated datasets.

By an~expensive distance computation we mean finding the~rotation matrix with Kearsley's method \cite{Kearsley1989} and calculating the~distance and derivatives with Equations (\ref{eq:dist}) and (\ref{eq:der_dist}). In the~original method, all distances are calculated expensively. In the~close structure method, the~majority of distances are just cheaply approximated with saved rotation matrices using Equations (\ref{eq:close_dist}) and (\ref{eq:close_der_dist}).

In the~original method, in a~usual step, we need $M$ (size of the~neighbourlist, or $M=0$ in case of no neighborlist) distance computations. In a~neighbourlist update step, every $L$ steps ($L=1$ in case of no neighborlist), we compute $N$ (number of reference structures) distances. On average, we need $M+\frac{N-M}{L}$ expensive distance computations per step.

In the~close structure method, in a~usual step, we need only one MSD computation, the~rest is approximated. Furthermore, in a~close structure reassignment step, we expensively recompute $N$ rotation matrices, so that we keep the~accuracy of distance computations for neighborlist updates. On average, we need $1+ \frac{N}{K}$ expensive distance computations per step, where the~close structure is reassigned (on average) every $K$ steps.

The~close structure method will accelerate metadynamics if $M+\frac{N-M}{L}>1+\frac{N}{K}$. From our experiments, if $\varepsilon$ is not too small, then $K \gg L$, as the~close structure was reassigned on average every few thousand steps for $\varepsilon = 0.01$\,nm$^2$ (10000 for cyclooctane, 3000 for Trp-cage) in the~100,000-step simulations we did. The~average number of reassignments depends on how often and how fast the~structure of molecule changes, so it varies with the~threshold $\varepsilon$, but also with the~length of the~simulation, the~starting structure, temperature and other simulation parameters. 

For the~simulations without the~neighbourlist, the~comparison is reduced to $N>1+\frac{N}{K}$ which clearly applies as $K \gg 1$. Therefore, the~number of distance computations is reduced approximately by an~order of $N$. 

For the~simulations with the~neighbourlist, the~specific values of $M$, $L$, $N$ and $K$ determine the~acceleration. It is safe to assume $M \ll N$ and $L \ll N$. If $K > L$, which is quite easily achievable for reasonable values of $\varepsilon$, then $\frac{N-M}{L}>\frac{N}{K}$. Thus, the~number of distance computations in simulations with the~NL can be reduced up to an~order of $M$. 

Naturally, the speed-up of distance computation accelerates the whole simulation only to some extent. This can be assessed using Ahmdal's law, see Equation~(\ref{eq:ahmdal}). The portion $p$ represents the percentage of time spent in functions we have accelerated, i.e. the distance computations, in the original implementation. The speed-up $s$ denotes the acceleration of this portion in our modified implementation. The whole computation can achieve only the speed-up $S$. For example, even an infinite speed-up of distance computation would accelerate the whole simulation only ten times, if the distance computations originally took up 90\% of computation time.

\begin{equation}
\label{eq:ahmdal}
 S = \frac{1}{(1-p) + \frac{p}{s}}
\end{equation}\normalsize{where $p$ is the~original portion of time spent in the~given function, $s$ is the~speed-up of the~accelerated function.}
\normalsize

More general performance behavior can be deduced from the analysis above. First, the acceleration brought by the close structure method decreases with the increased cost of molecular dynamics due to Ahmdal's law. This applies especially in the case of larger molecules. The cost of computing MSD with the original method also grows with the number of atoms, but to a much lesser extent than the demands of MD. Therefore, a higher acceleration is expected with smaller molecules. With larger molecules, the best we can achieve is to get closer to the performance of molecular dynamics without metadynamics. Second, the acceleration brought by the close structure method increases with the decreasing stride of neighborlist's update ($L$), but does not depend much on its size. Therefore, a higher acceleration is expected in the case of no neighborlist. And finally, the acceleration brought by the close structure method decreases with the increased number of reference structures, especially with the neighborlist updated 
often.
    
\subsection{Theoretical Speedup for Evaluated Datasets}
We evaluated the theoretical speed-up expected in simulations with our datasets. Moreover, we assessed the speed-up of the whole simulation using Ahmdal's law and information from profiling.

For cyclooctane, we need 512 MSD computations each step with the~original method and an~average of $1+512/10000 = 1.05$ with the~close structure method. The~theoretical speed-up for MSD computation reaches $512/1.05 = 488$. 

For Trp-cage, we need on average $50 + (2120-50)/50 = 91.4$ MSD computations with the~original method and $1+2120/3000 = 1.7$ with the~close structure method. The~theoretical speed-up for MSD computation reaches $91.4/1.7 = 54$.

The~theoretical speed-up for the~whole simulation can be assessed using Ahmdal's law and information from profiling. We combined the~information from log files (time spent with whole metadynamics and evaluation of collective variables) and from Intel VTune Profiler (time spent computing the~distance) to assess the~percentage of computation time spent in various functions in cyclooctane and Trp-cage simulations with one MPI process and one OpenMP thread, see Table~\ref{tab:ahmdal}. Metadynamics (MTD) includes the~evaluation of collective variables (CV), initialization, logging and addition of Gaussian hills. CV evaluation requires distance computations (MSD) before the calculation of its own value. MSD computation consists of matrix diagonalization (DSYEVR), the calculation of the distance and the derivatives. The~speed-up $S_{MSD}$ represents the~maximal theoretical speed-up of the~whole simulation according to Ahmdal's law (see Equation~(\ref{eq:ahmdal})) taking into account the speed-up of MSD computation as 
assessed above.

\begin{table}[h]
\centering
\caption{Time portions spent in various metadynamics functions and the~maximal speed-up of the~whole simulation} \label{tab:ahmdal}
 \begin{tabular}{m{5em}rrrrrr}
 \hline
  &  MTD & CV & MSD & DSYEVR & $S_{MSD}$\\
  \hline
  cyclooctane & 99.7\%  & 98.5\% & 93\% & 70\% & 14\\
  Trp-cage & 53\%  & 50\% & 43\% &  23\% & 1.7 \\
  \hline
 \end{tabular}
\end{table}


As expected, cyclooctane simulation would greatly benefit from the~acceleration due to a smaller size and no neighborlist. Trp-cage simulation can achieve only modest improvement as the most of the computational cost lies in MD code.

\subsection{Practical Speedup and Scalability}
For the~practical evaluation of speed-up and scalability, we ran simulations without metadynamics (Gromacs), with metadynamics and the~original method for MSD computation (Gromacs+Plumed) and with metadynamics modified with the~close structure method (Gromacs+modified Plumed). Furthermore, we executed them for various numbers of MPI processes and OpenMP threads. As a~speed measure, we took \emph{ns/day} stated by Gromacs in its logs. As a~reference point for the~speed-up, we considered Gromacs with original Plumed ran with one MPI process and one OpenMP thread.

Out of four runs for each variant and MPI/OpenMP configuration, we considered the~minimal running time, i.e. the~highest speed, to eliminate the~random interference of the~operating system's background activity. 
\begin{figure}[h!]
\centering
 \includegraphics[width=0.8\hsize]{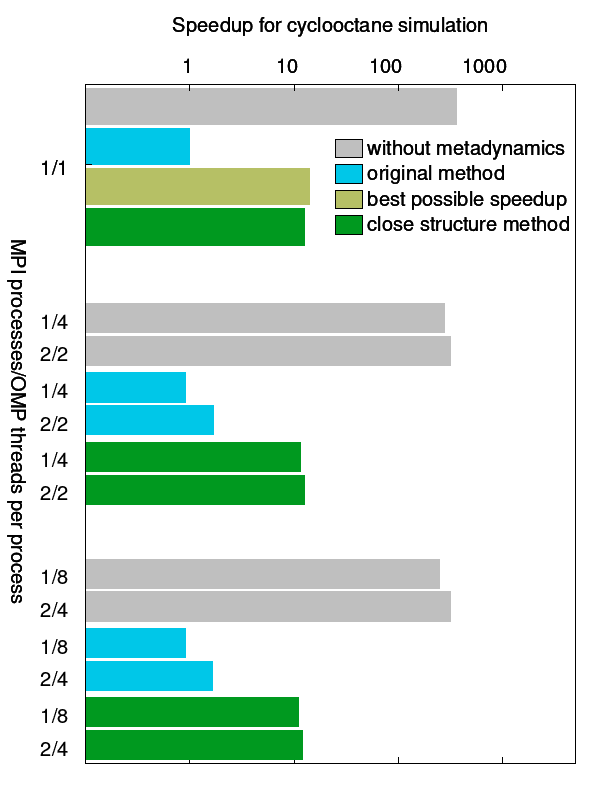}
 \caption{Speedup and strong scalability of cyclooctane simulations. Notice the~logarithmic scale of x axis, the~reference point is the~blue bar with the~original method. The~grey bar with molecular dynamics without metadynamics puts into perspective how metadynamics increases the~cost.} \label{fig:scalability_cyclooctane}
\end{figure}

\begin{figure}[h!]
\centering
 \includegraphics[width=0.8\hsize]{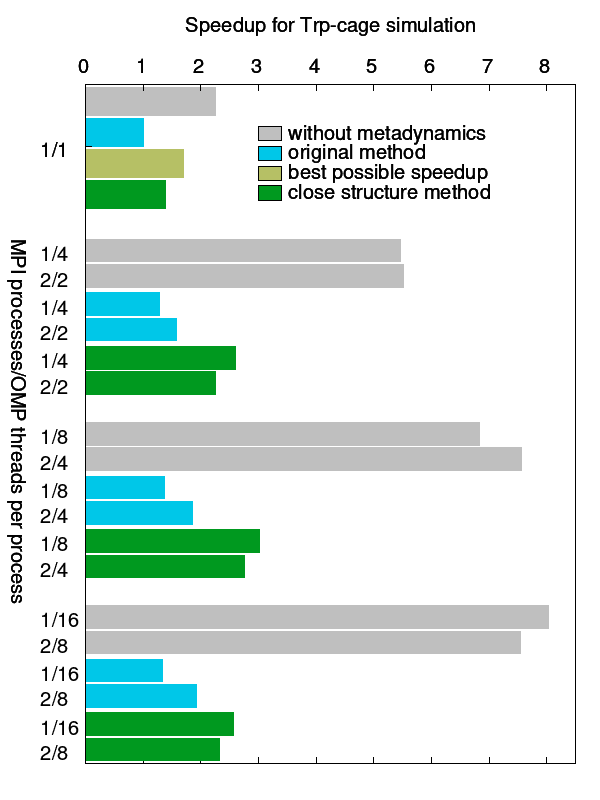}
 \caption{Speedup and strong scalability of the~Trp-cage simulations. The~reference point is the~blue bar with the~original method. The~grey bar with molecular dynamics without metadynamics puts into perspective how metadynamics increases the~cost.} \label{fig:scalability_trpcage}
\end{figure}

The~upper parts of Figures~\ref{fig:scalability_cyclooctane} and \ref{fig:scalability_trpcage} show the~speed-ups for the~basic case of one MPI process and one OpenMP thread. For both molecular systems, the~close structure method almost reaches the~maximal theoretical speed-up as calculated in the~previous subsection. 

Cyclooctane simulations with the~close structure method achieved a~speed-up of an~order of magnitude, notice the~logarithmic scale on the x-axis of Figure~\ref{fig:scalability_cyclooctane}. The~close structure method almost reached to the~maximal theoretical performance, albeit it still lags behind Gromacs without metadynamics. That is caused by Plumed overhead like initialization, logging, Gaussian hills addition and CV evaluation. 

Trp-cage simulations with the~close structure method achieved a~more modest speed-up due to the~major cost of molecular dynamics. Furthermore, the practical and theoretical speed-up differ more than in cyclooctane simulation. We suggest it is caused by a smaller portion of time spent in matrix diagonalization function, as shown in Table \ref{tab:ahmdal}. The distance computation takes 43\% of computation time. Out of that, only half is spent in diagonalization, as opposed to three-quarters in the case of cyclooctane. The rest is spent in the computation of distance and especially derivatives, which scales with the size of the molecule. We reduced the number of calls to DSYEVR with the close structure method, but the other half of computation remained unchanged: both Gromacs+Plumed and Gromacs+modified Plumed 
contain the same code optimizations. Nevertheless, we got much closer to the performance of Gromacs alone, thus gaining the advantage of enhanced sampling for only a small cost.  

The~lower parts of Figures~\ref{fig:scalability_cyclooctane} and \ref{fig:scalability_trpcage} show the~speed-ups of strong scaling for various combinations of MPI processes and OpenMP threads. 

First, compare the~performances with the~same total number of threads, i.e. 1/4 with 2/2, 1/8 with 2/4 and 1/16 with 2/8. In all cases for the~original method, two MPI processes perform better than one with twice as many threads. The~reason for that stems from the~implementation of CV calculation in Plumed. The~loop over reference structures in the~neighbourlist (line 2 in Algorithm~\ref{alg:model}) is parallelized only with MPI. As the~close structure method accelerates the~computation within that loop, the~second MPI process does not bring any advantage. In the~case of Trp-cage, it even slows down the~computation due to the~communication and synchronization cost. Decreasing speed with more resources (8 total threads for cyclooctane, 16 for Trp-cage) suggests that the~overhead caused by parallelization exceeds its benefits. 

Second, compare performances for one MPI process, i.e. 1/4, 1/8 and 1/16. Any speed-up here can come only from faster molecular dynamics. For cyclooctane simulations, a~small amount of computation saturates even a~single core, thus performance stays about the~same for all three variants. For Trp-cage simulation, some improvement appears as molecular dynamics requires more computation due to the~protein's size. The simulations with the close structure method also exhibit this trend.

Overall, the~close structure method reduces the~gain of another MPI process but does not interfere with the~general scalability behavior.

\section{Conclusion}
In this work, we address high computational demands of distance calculations in path/property map collective variables in metadynamics. The~original method expensively evaluates the~rotation matrices needed to superimpose the~current structure to the~reference structures in each step for many reference structures. By introducing a~\emph{close structure}, we reuse those matrices in next steps and cheaply approximate the~vast majority of distance computations. 

We thoroughly evaluated accuracy elsewhere \cite{Pazurikova2017} and concluded that simulations with the~close structure method explore the~same energy landscape as simulations with the~original method, despite different trajectories.

Here, we presented the tuned implementation of our method ready for production use and evaluated its performance. We calculated the~number of distance computations for both the~original and close structure method and assessed the~theoretical speed-up of the~whole simulation. The~number of expensive MSD computations decreased approximately by the~order of the~neighbourlist size or the~number of all the~reference structures in case of no neighborlist. Clearly, the~accelerated distance computation is expressed in the~speed-up of the~whole simulation only to the~extent proportional to the~time spent on distance computation in the~original method. That depends on the~neighborlist size and update stride, but mostly on the~cost of the~molecular dynamics part, determined by the~molecule's size.

The~experiments showed that the~actual speed-up closely approaches the~theoretical one for both evaluated molecular systems. Apart from the~close structure approximation, minor code optimizations in distance computation contributed to this, including enforcing of loop vectorization and inlining of frequently called functions. These optimizations did not help the~original method, as it spent the~majority of time in the~finely-tuned BLAS routine. However, unvectorized loops became prominent after the~employment of the~close structure, so even minor code optimizations pushed the~practical speed-up further.

The~close structure method affects the~scalability of Plumed code, diminishing the~gain of an~additional MPI process. This was expected as we reduced the~amount of computation so that even one core suffices.

Overall, the positive effects of the close structure approximation are prominent especially in smaller molecules with frequently updated neighborlist or no neighborlist at all. With larger molecules, the method diminishes the overhead of metadynamics and approaches the performance of sole molecular dynamics simulation with the advantage of enhanced sampling. The error brought by the approximation of the close structure method does not negatively influence the simulation, the explored landscapes remain the same.

Here we presented a~typical application of the~close structure method, molecular dynamics with metadynamics using property map collective variables to calculate the~bias potential. The~close structure method would benefit any CV-based method of enhanced sampling, provided it uses mean square distance as a~measure. Our method significantly reduces the~overhead of metadynamics in simulations, getting closer to the~ideal situation: molecular dynamics with the~advantage of enhanced sampling---but without significant additional cost. 

 \section*{Acknowledgments}
 
This work was supported by Czech Science Foundation (15-17269S) and LM2015047 Czech National Infrastructure for Biological Data. Computational resources were provided by the~CESNET LM2015042 and the~CERIT Scientific Cloud LM2015085, provided under the~programme ``Projects of Large Research, Development, and Innovations Infrastructures''.
 





\bibliographystyle{elsarticle-num}








\end{document}